# Near-UV and optical observations of the transiting exoplanet TrES-3b

Jake D. Turner,[1,2]★ Brianna M. Smart,[2] Kevin K. Hardegree-Ullman,[2]
Timothy M. Carleton,[2] Amanda M. Walker-LaFollette,[2] Benjamin E. Crawford,[2]
Carter-Thaxton W. Smith,[2] Allison M. McGraw,[2] Lindsay C. Small,[2]
Marco Rocchetto,[3] Kathryn I. Cunningham,[4] Allison P. M. Towner,[2] Robert Zellem,[1]
Amy N. Robertson,[2] Blythe C. Guvenen,[2] Kamber R. Schwarz,[2]
Emily E. Hardegree-Ullman,[5] Daniel Collura,[5] Triana N. Henz,[2] Cassandra Lejoly,[2]
Logan L. Richardson,[2] Michael A. Weinand,[6] Joanna M. Taylor,[7]
Michael J. Daugherty,[2] Ashley A. Wilson[6] and Carmen L. Austin[2]

[1]Lunar and Planetary Laboratory, University of Arizona, Tucson, AZ 85721, USA
[2]Steward Observatory, University of Arizona, Tucson, AZ 85721, USA
[3]Department of Physics & Astronomy, University College London, Gower Street, London WC1E 6BT
[4]Department of Computer Science, University of Arizona, Tucson, AZ 85721, USA
[5]Department of Physics, Applied Physics and Astronomy, Rensselaer Polytechnic Institute, Troy, NY 12180, USA
[6]Department of Physiology, University of Arizona, Tucson, AZ 85724, USA
[7]Department of Astronomy, Indiana University, Bloomington, IN 47405, USA

## ABSTRACT

We observed nine primary transits of the hot Jupiter TrES-3b in several optical and near-UV photometric bands from 2009 June to 2012 April in an attempt to detect its magnetic field. Vidotto, Jardine and Helling suggest that the magnetic field of TrES-3b can be constrained if its near-UV light curve shows an early ingress compared to its optical light curve, while its egress remains unaffected. Predicted magnetic field strengths of Jupiter-like planets should range between 8 G and 30 G. Using these magnetic field values and an assumed $B_*$ of 100 G, the Vidotto et al. method predicts a timing difference of 5–11 min. We did not detect an early ingress in our three nights of near-UV observations, despite an average cadence of 68 s and an average photometric precision of 3.7 mmag. However, we determined an upper limit of TrES-3b's magnetic field strength to range between 0.013 and 1.3 G (for a 1–100 G magnetic field strength range for the host star, TrES-3) using a timing difference of 138 s derived from the Nyquist–Shannon sampling theorem. To verify our results of an abnormally small magnetic field strength for TrES-3b and to further constrain the techniques of Vidotto et al., we propose future observations of TrES-3b with other platforms capable of achieving a shorter near-UV cadence. We also present a refinement of the physical parameters of TrES-3b, an updated ephemeris and its first published near-UV light curve. We find that the near-UV planetary radius of $R_p = 1.386^{+0.248}_{-0.144}\ R_{\rm Jup}$ is consistent with the planet's optical radius.

## 1 INTRODUCTION

TrES-3b, one of ∼282[1] confirmed transiting exoplanets, is a hot Jupiter orbiting a G-type star with a short orbital period of

1.306 19 ± 0.000 01 d, a mass of 1.92 ± 0.23 $M_{\rm Jup}$ and a radius of 1.295 ± 0.081 $R_{\rm Jup}$ (O'Donovan et al. 2007). Several follow-up primary transit photometric studies have confirmed these planetary parameters (e.g. Gibson et al. 2009; Sozzetti et al. 2009; Colón et al. 2010; Ballard et al. 2011; Christiansen et al. 2011; Lee et al. 2011). Christiansen et al. (2011) saw long-term variability over a 12 d span in TrES-3's light curve, which they attributed to starspots. The observations by Christiansen et al. (2011) were not able to further constrain TrES-3's rotation period because O'Donovan et al. (2007) found it to be >21 d. In addition, Sozzetti et al. (2009) found that there may be a deviation from a constant period in TrES-3b caused by other orbiting bodies in the system. However, Gibson et al. (2009) ruled out any additional planets in the interior and exterior 2:1 resonances. Multiple studies (de Mooij & Snellen 2009; Sada et al. 2012) have shown that the near-infrared (near-IR) planetary radius of TrES-3b is consistent with the optical planetary radius (O'Donovan et al. 2007). Secondary eclipse measurements in the optical and near-IR, conducted by Winn et al. (2008), de Mooij & Snellen (2009), Croll et al. (2010), Fressin, Knutson & Charbonneau (2010) and Christiansen et al. (2011), found that TrES-3b has efficient re-circulation and no temperature inversion in its upper atmosphere.

The transit method for detecting and observing exoplanets is the only method that allows direct measurements of the planetary radius and the determination of planetary characteristics such as average density, surface gravity, atmospheric composition, semi-major axis and eccentricity (Charbonneau et al. 2007). When combined with radial velocity measurements, transit measurements can further constrain the mass of the planet (Charbonneau et al. 2007). In addition, Lazio et al. (2009) theorized that the transit method can be used to detect exoplanet magnetic fields.

All the gas giant planets in our Solar system possess magnetic fields (Russell & Dougherty 2010) and it is expected through interior structure models that extrasolar gas giants should also have magnetic fields (Sánchez-Lavega 2004). Detecting and studying the magnetic fields of exoplanets will allow for the investigation of many more properties of exoplanets, including interior structure and rotation period (Lazio et al. 2009), the presence of extrasolar moons [e.g. the modulations in Jupiter's magnetic field can be contributed to the presence of Io (Lazio et al. 2009)] and atmospheric retention (Lazio et al. 2009; Cohen et al. 2011). Furthermore, Grießmeier et al. (2005) suggest that the magnetic field of the Earth helps contributing to its habitability by deflecting cosmic rays and stellar wind particles; exoplanets could also exhibit this characteristic (Lazio et al. 2009). Studying the magnetic fields of hot Jupiters will help lay the foundation for the characterization of magnetic fields around Earth-like planets and consequently, it will aid in the search for life outside our Solar system.

Several methods have been used to attempt to detect magnetic fields of exoplanets. Farrell, Desch & Zarka (1999), Zarka et al. (2001), Bastian, Dulk & Leblanc (2000) and Lazio et al. (2010a) suggest that the most direct method for detecting the magnetic field of an exoplanet is through radio emission from the planet generated by electron–cyclotron maser interactions. Specifically, this electron–cyclotron maser radio emission is caused by currents within the planet's magnetosphere formed through interactions between the solar wind and the magnetosphere being directed into the planet's magnetic polar regions. However, many studies conducted to find exoplanet radio emission resulted in non-detections (e.g. Yantis, Sullivan & Erickson 1977; Winglee, Dulk & Bastian 1986; Bastian et al. 2000; Ryabov, Zarka & Ryabov 2004; George & Stevens 2007; Lazio & Farrell 2007; Smith et al. 2009; Lecavelier Des Etangs et al. 2009, 2011; Lazio et al. 2010a, b). Alternatively, Cuntz, Saar & Musielak (2000), Saar & Cuntz (2001) and Ip, Kopp & Hu (2004) proposed that the interaction of the magnetic field of an exoplanet and its host star could produce detectable changes in the star's outer layers and corona in phase with the planet's orbit. This indirect method of detecting the magnetic field of an exoplanet was validated through observations by Shkolnik, Walker & Bohlender (2003),

Shkolnik et al. (2005, 2008) and Gurdemir, Redfield & Cuntz (2012).

In this paper, we use another method to attempt to detect the magnetic field of an exoplanet, described by Vidotto, Jardine & Helling (2010), Vidotto et al. (2011a), Vidotto, Jardine & Helling (2011b), Vidotto et al. (2011c), Lai, Helling & van den Heuvel (2010) and Llama et al. (2011). They suggest that a transiting exoplanet with a magnetic field will show an earlier transit ingress in the near-ultraviolet (near-UV) wavelengths than in the optical wavelengths, while the transit egress times would be the same. These authors explained this effect by the presence of a bow shock in front of the planet formed through interactions between the stellar coronal material and the planet's magnetosphere. Furthermore, if the shocked material is sufficiently optically thick, it will absorb starlight and cause an early ingress in the near-UV light curve (Vidotto et al. 2011b, see fig. 6). The difference between ingress times in different wavelength bands can be used to constrain the properties of the planet's magnetic field. An early near-UV ingress has been observed in one transiting exoplanet, WASP-12b (Fossati, Haswell & Froning 2010, hereafter FHF10). Observations by FHF10 of WASP-12b using the *Hubble Space Telescope* (*HST*) with the NUVA (253.9–258.0 nm) near-UV filter indicate that the near-UV transit started approximately 25–30 mins earlier than its optical transit. The spectral region covered by the NUVA filter includes strong resonance lines from Na I, Al I, Sc II, Mn II, Fe I, Co I and Mg I (Morton 1991, 2000) and according to FHF10, these spectral lines likely caused the deeper transit in WASP-12b. Using these observations, Vidotto et al. (2010) determined an upper limit for the magnetic field of WASP-12b to be ∼24 G. Furthermore, Vidotto et al. (2011a, hereafter VJH11a) predicted that near-UV ingress asymmetries should be common in transiting exoplanets. However, VJH11a do not specify whether this effect can be seen only in narrow-band spectroscopy [as with the WASP-12b observations (FHF10)] or broad-band near-UV photometry. In addition, Vidotto et al. (2011c) predicted that bow shock variations should be common and are caused by eccentric planetary orbits, azimuthal variations in coronal material (unless the planet is in the corotation radius of the star) and time-dependent stellar magnetic fields (e.g. coronal mass ejections, magnetic cycles, stellar wind changes). Consequently, the near-UV light curve of exoplanets predicted by VJH11a will exhibit temporal variations. For example, Haswell et al. (in preparation) saw their near-UV light curve of WASP-12b start earlier than FHF10's near-UV observations.

We chose TrES-3b for our study because it is listed as one of the top ten candidates predicted by VJH11a to exhibit near-UV asymmetries. In addition, the WASP-12 and TrES-3 systems have very similar physical characteristics (see Table 1 for a summary). Therefore, since FHF10 observed near-UV asymmetries in WASP-12b it would be reasonable to assume that TrES-3b could also exhibit this effect.

If observed, a difference in timing between the near-UV and optical light curves of TrES-3b can be used to determine the planetary magnetic field, $B_p$, with the following equation derived from Vidotto et al. (2011b):

$$B_p = B_* \left( \frac{R_*}{a R_p} \right)^3$$
$$\times \left\{ \frac{2\delta t}{t_d} \left[ \left( R_*^2 - \left\{ \frac{a \cos i}{R_*} \right\}^2 \right)^{1/2} + R_p \right] + R_p \right\}^3, \quad (1)$$

where $B_*$ is the host star's magnetic field, $R_*$ is the host star's radius, $a$ is the semi-major axis, $R_p$ is the planet radius, $\delta t$ is the difference in timing between the near-UV and optical ingress, $t_d$ is the optical transit duration and $i$ is the orbital inclination. The parameters $R_p$, $a$, $t_d$ and $i$ can all be derived from the optical light curve, and $B_*$ and $R_*$ can be determined from previous studies (e.g. O'Donovan et al. 2007, Sozzetti et al. 2009) of the TrES-3 host star. VJH11a predicts a minimum planetary magnetic field required to sustain a magnetosphere for TrES-3b to be $B_p^{min} = 0.0047\,B_*$. Using equation (1), we derived a minimum timing difference between the near-UV and optical ingress times of 3 s.

**Table 1.** Comparison of the TrES-3 and WASP-12 systems.

| Planet Name | $M_b$ ($M_{Jup}$) | $R_b$ ($R_{Jup}$) | $P_b$ (d) | $a$ (au) | Spec. Type | $M_*$ ($M_\odot$) | $R_*$ ($R_\odot$) | [Fe/H] | $B_p/B_*^a$ (per cent) | $\delta t^b$ (s) |
|---|---|---|---|---|---|---|---|---|---|---|
| TrES-3b[1] | 1.92 | 1.31 | 1.31 | 0.023 | G | 0.90 | 0.80 | −0.19 | 0.47 | 3 |
| WASP-12b[2] | 1.41 | 1.79 | 1.09 | 0.023 | G0 | 1.35 | 1.57 | 0.30 | 3.2 | 5 |

[1]Reference for $M_b$, $R_b$, $P_b$, $a$, Spec Type, $M_*$, $R_*$; O'Donovan et al. (2007). Reference for [Fe/H]; Sozzetti et al. (2009).

[2]Reference for $M_b$, $R_b$, $P_b$, $a$, Spec Type, $M_*$, $R_*$, [Fe/H]; Hebb et al. (2009).

$^a$$B_p/B_*$ is the minimum planetary magnetic field relative to the stellar one that is required to sustain a magnetosphere. Reference for $B_p/B_*$; Vidotto et al. (2011a).

$^b$$\delta t$ is the minimum timing difference between the optical and near-UV ingress times calculated from equation (1) inputting $B_p/B_*$. Reference for $\delta t$; Vidotto et al. (2011a).

The magnetic field strength of Jupiter-like exoplanets have been predicted based on several different scaling laws (see Christensen 2010 for a summary). Reiners & Christensen (2010) found that the energy flux (correlated with its age and mass) of a planet controls the magnetic field strength only if the rotation period of the object is above a certain critical limit. These authors predicted that for a 1 $M_{Jup}$ exoplanet its magnetic field strength will range from ∼8 G to ∼30 G assuming an age of 0.1 and 3.7 Gyr (TrES-3's age is $0.9^{+2.8}_{-0.8}$ Gyr; Sozzetti et al. 2009), respectively. Alternatively, Sánchez-Lavega (2004) used the rotation period and planetary mass to predict an upper limit magnetic field strength of a hot Jupiter (with an orbital period of 10–20 h) to be ∼30 G, assuming spin–orbit synchronism. The upper limit by Reiners & Christensen (2010) is consistent with the values predicted by Sánchez-Lavega (2004). TrES-3b's magnetic field strength could be slightly lower than 8–30 G depending on the amount of tidal braking it experiences (Reiners & Christensen 2010). Assuming an 8–30 G magnetic field strength range for TrES-3b and a reasonable maximum magnetic field strength for its host star based off of previous studies of G-type stars (Reiners 2012; Plachinda 2004; Plachinda & Tarasova 2000, 1999) of 100 G, we find a timing difference between 5 and 11 min (a lower stellar magnetic field would increase this timing difference) using equation (1). This timing resolution is well within reach for ground-based metre-sized telescopes, like the Steward Observatory 1.55-m Kuiper Telescope used for the near-UV observations in this study.

The goal of this paper is to determine if ground-based observations of TrES-3b transits in broad-band near-UV and optical wavelengths are capable of constraining the planet's magnetic field. Additionally, using our data set, we update the planetary system parameters, search for a wavelength dependence in the planetary radius and present a new ephemeris to help with future observations.

## 2 OBSERVATIONS AND DATA REDUCTION

Most of our observations were conducted at the Steward Observatory 1.55-m Kuiper Telescope on Mt. Bigelow near Tuc-

son, Arizona, using the Mont4k CCD. The Mont4k CCD contains a 4096 × 4096 pixel sensor with a field of view (FOV) of 9.7arcmin × 9.7 arcmin. We used 3 × 3 binning to achieve a resolution of 0.43 arcsec pixel$^{-1}$ and a 4096 × 2048 pixel subframe with a FOV of 9.7arcmin × 4.85 arcmin, to shorten read-out time to ∼10 s. Our observations were taken with the Bessell U (303–417 nm), Harris B (330–550 nm), Harris V (473–686 nm) and Harris R (550–900 nm) photometric band filters. Specifically, the Bessell U filter is a near-UV filter with a transmission peak of 70 per cent near 370 nm. To ensure accurate time keeping, an onboard system clock was automatically synchronized with GPS every few seconds throughout the observational period. Two of our observations were conducted at the University of London Observatory (ULO) 0.36-m EAST and WEST Celestron CGE 1400 telescopes equipped with SBIG STL-6303E CCD sensors using the Johnson–Cousins R (523–940 nm) filter. The SBIG STL-6303E CCDs contain 3072 × 2048 pixel sensors with a FOV of 18 arcmin × 24 arcmin and 0.86 arcsec pixel$^{-1}$ resolution. To ensure accurate time in these observations, the clocks were synchronized with a Network Time Protocol server. The ULO 0.36-m telescopes have provided exceptional transit results in previous studies (e.g. Fossey, Waldmann & Kipping 2009). Due to excellent autoguiding, there was no more than a 2 pixel (∼1.29arcsec) and 4 pixel (∼3.44arcsec) drift in the centroid of TrES-3 in all our data sets from the 1.55-m Kuiper and the 0.36-m EAST and WEST ULO telescopes, respectively. It is important to note that for the2011 October 14 transit, we experienced technical difficulties with the Mont4k CCD from 15 min before the transit ingress to the expected transit ingress. Therefore, we were only able to obtain four images during that time period and this resulted in a non-uniform sampling for the 2011 October 14 transit. Seeing ranged from 1.49–3.60 arcsec throughout our complete set of observations. A summary of all our observations is displayed in Table 2. The Out-of-Transit (OoT) baseline in all transits achieved a photometric root-mean-squared (RMS) value between 1 and 4 mmag, which are typical values for high signal-to-noise (S/N) ratio transit photometry with a median RMS of 2 mmag for both the Mont4k on the 1.55-m Kuiper telescope (Dittmann et al. 2009, 2010a, b, 2012; Scuderi et al. 2010) and ULO's 0.36-m EAST and WEST telescopes (Fossey et al. 2009).

Using standard IRAF[2] techniques, each of our images was bias-subtracted and flat-fielded. In our calibration process for the 2009 July 22 observations, we tested to see if we obtained better S/N ratios by calibrating our data with different numbers of flat-field images (flats). We performed the calibration steps on our data set with 5, 10, 20, 50, 100 and 200 flats. In order to quantify the effect of using different numbers of flats, we performed an analysis of variance (ANOVA) test on the average S/N ratio for our target star and five reference stars. The ANOVA test is designed to check whether our null hypothesis (that the means of our sets of flat-field data are statistically the same) can be accepted or rejected based upon the *p*-value. The *p*-value is the statistical result of an ANOVA test that measures the probability of obtaining a result significantly different from what was actually observed. A *p*-value greater than 0.5 implies that the null hypothesis sufficiently describes the set of data. From our analysis, we find a *p*-value of 1, indicating that we can accept the null hypothesis and that there is no significant difference between the number of flats we use to adequately reduce the noise in our images. Therefore, to optimize telescope time, we obtained and used 10 flats in all subsequent observations and reductions.

[2] IRAF is distributed by the National Optical Astronomy Observatory, which is operated by the Association of Universities for Research in Astronomy, Inc., under cooperative agreement with the National Science Foundation.

**Table 2.** Journal of observations.

| Date (UT) | Filter[1] | Telescope[2] | Cadence (s) | OoT RMS[3] (mmag) | Res RMS[4] (mmag) | Red noise[5] (mmag) | Seeing (arcsec) |
|---|---|---|---|---|---|---|---|
| 2009 June 13 | B | Kuiper | 34 | 2.18 | 2.83 | $9.4^{+4.3}_{-4.2}$ | 1.50–1.84 |
| 2009 June 22 | R | Kuiper | 21 | 1.53 | 1.65 | $4.1^{+2.4}_{-2.3}$ | 1.49–1.60 |
| 2009 July 04 | V | Kuiper | 30 | 2.83 | 2.57 | $5.6^{+3.4}_{-3.5}$ | 1.56–2.30 |
| 2011 October 14 | U | Kuiper | 67 | 4.38 | 3.70 | $13^{+8}_{-8}$ | 1.77–2.58 |
| 2011 November 04 | U | Kuiper | 67 | 3.92 | 3.08 | $15^{+11}_{-8}$ | 1.56–1.92 |
| 2012 March 25 | U | Kuiper | 69 | 3.03 | 3.13 | $11^{+4}_{-3}$ | 1.57–1.88 |
| 2011 March 28(1) | J–C R | ULO EAST | 130 | 2.56 | 1.86 | $4.9^{+4.4}_{-4.3}$ | 2.80–3.60 |
| 2011 March 28(2) | J–C R | ULO WEST | 149 | 2.60 | 1.92 | $6.8^{+5.6}_{-4.4}$ | 2.80–3.60 |
| 2012 April 11 | B | Kuiper | 47 | 2.17 | 2.25 | $10^{+3}_{-2}$ | 2.24–2.49 |

[1]Filter used: *B* is the Harris B (330–550 nm), *R* is the Harris R (550–900 nm), *V* is the Harris V (473–686 nm), *U* is the Bessell U (303–417 nm) and *J-C R* is the Johnson–Cousins R (523–940 nm).

[2]Telescope used: Kuiper is the 1.55-m Kuiper Telescope, ULO EAST is the ULO 0.36-m EAST telescope and ULO WEST is the ULO 0.36-m WEST telescope.

[3]OoT RMS relative flux.

[4]Residual (res) RMS flux after subtracting the TAP best-fitting model from the data.

[5]Red noise (temporally correlated noise) calculated from TAP.

To produce the TrES-3b light curves, we performed aperture photometry with the PHOT task in the IRAF DAOPHOT package. We created light curves for TrES-3b and five reference stars using three different aperture radii. We used a different set of aperture radii depending on the conditions of each observing night. A synthetic light curve was produced by averaging the light curves from our reference stars, and the final light curves of TrES-3b were normalized by dividing by this synthetic light curve. Several different combinations of reference stars and aperture radii were considered, and we picked the combination that produced the lowest scatter in the OoT data points. The unbinned light curves for all our data are shown in Fig. 1.

## 3 LIGHT CURVE ANALYSIS

### 3.1 Light curve modelling

We modelled our transit light curves with two different publicly available modelling software packages: the Transit Analysis Package[3] (TAP; Mandel & Agol 2002; Carter & Winn 2009; Gazak, Johnson & Tonry 2011; Eastman, Agol & Gaudi 2012) and

JKTEBOP[4] (Southworth, Maxted & Smalley 2004a, b). TAP utilizes Markov Chain Monte Carlo techniques to fit transit light curves with a standard Mandel & Agol (2002) model and estimates parameter uncertainties with a wavelet likelihood function (Carter & Winn 2009, hereafter CW09). Using wavelet likelihood techniques to estimate uncertainties are more reliable than $\chi^2$ likelihood techniques because they account for parameters affected from temporally uncorrelated and correlated noise (CW09; Johnson et al. 2012). A more detailed explanation of the wavelet likelihood technique can be found in CW09. JKTEBOP was originally developed from the EBOP program written for eclipsing binary star systems (Etzel 1981; Popper & Etzel 1981) and implements the Nelson–Davis–Etzel eclipsing binary model (Nelson & Davis 1972). One advantage of JKTEBOP over TAP is that it uses biaxial spheroids to model the object of interest (in our case a planet), allowing for departures from sphericity (Southworth 2010). This feature of JKTEBOP is important in our analysis because if TrES-3b exhibits an early near-UV ingress, the planet would appear to be non-spherical (e.g. Llama et al. 2011, see fig. 2). In addition, JKTEBOP uses the Levenberg–Marquadt Monte Carlo technique to compute errors (Press et al. 1992; Southworth 2010; Hoyer et al. 2012). We only used JKTEBOP to search for any non-spherical asymmetries between the near-UV and optical light curves (see Section 4.1) and used TAP for a majority of the analysis of our data set.

Using TAP, we modelled each transit individually using five chains with lengths of 40 000 links each. During analysis, the inclination (*i*), time of mid-transit ($T_c$), planet-to-star radius ratio ($\frac{R_p}{R_*}$) and scaled semi-major axis ($\frac{a}{R_*}$) were allowed to float. Eccentricity (*e*), argument of periastron ($\omega$), the quadratic limb darkening coefficients ($\mu_1$ and $\mu_2$) and the orbital period ($P_b$) of the planet were fixed. Both *e* and $\omega$ were set to zero (suggested by O'Donovan et al. 2007; Fressin et al. 2010) and the period was set to $P_b = 1.306186$ d (Christiansen et al. 2011). The linear ($\mu_1$) and quadratic ($\mu_2$) limb darkening coefficients in each respective band were taken from Claret (2000) and interpolated to the effective temperature ($T_{eff}$), metallicity ([Fe/H]) and surface gravity (log *g*) of TrES-3 taken from

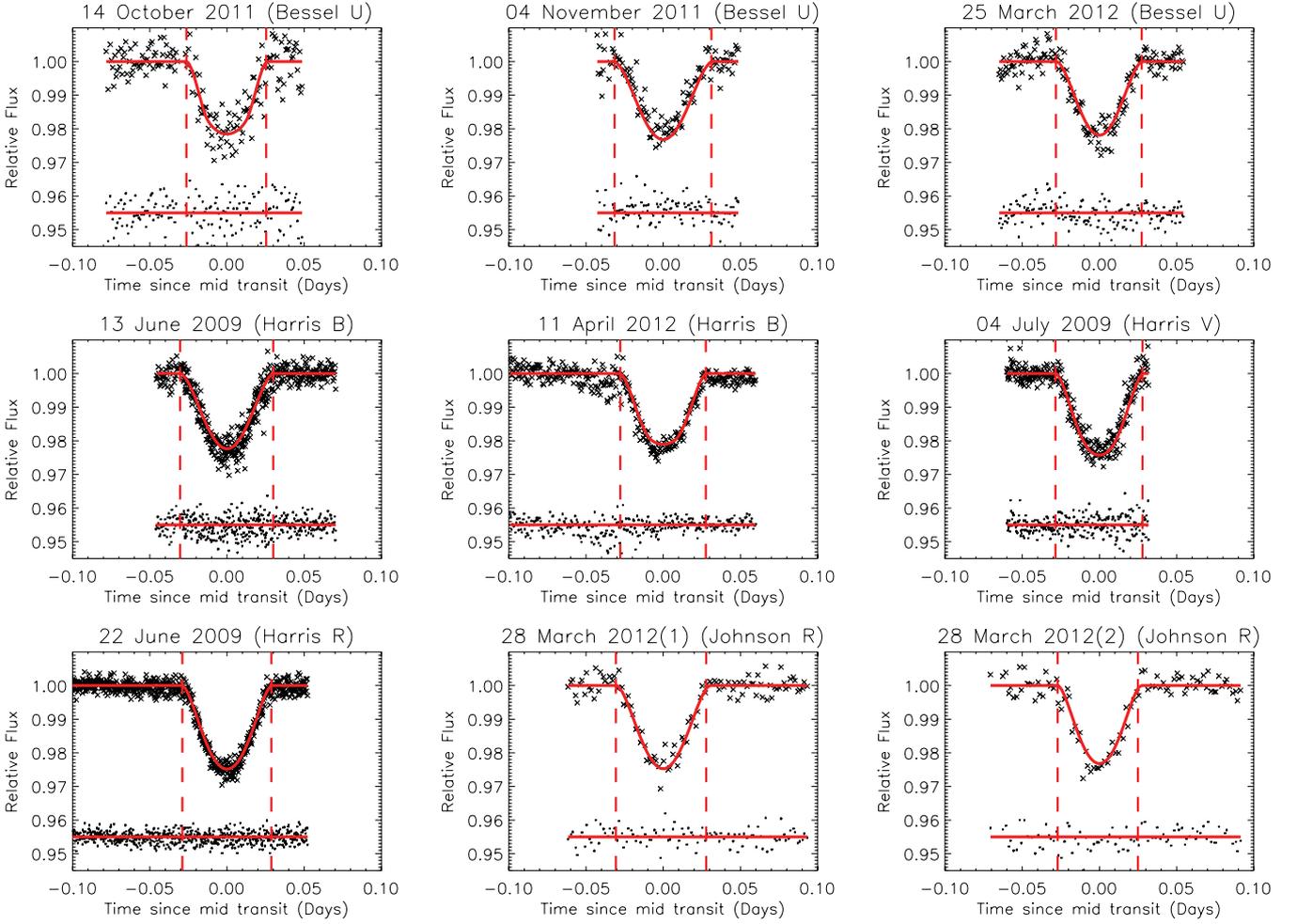

**Figure 1.** Light curves of TrES-3b for each observation. The best-fitting models obtained from TAP are shown as solid red lines. The residuals are shown below each transit light curve. The TAP best-fitting model predicted ingress and egress points are shown as dashed-red vertical lines. See Table 2 for the cadence, OoT RMS flux, residual RMS flux and red (temporally correlated) noise of each light curve.

**Table 3.** Limb darkening coefficients adopted for TrES-3b[1].

| Filter | Linear coefficient ($\mu_1$) | Quadratic coefficient ($\mu_2$) |
|---|---|---|
| Bessell U | 0.818 76 | 0.045 02 |
| Harris B | 0.637 12 | 0.179 94 |
| Harris V | 0.438 80 | 0.292 64 |
| Harris R | 0.341 56 | 0.318 18 |
| Johnson–Cousins R | 0.341 56 | 0.318 18 |

[1] $\mu_1$ and $\mu_2$ were taken from Claret (2000) for each band and interpolated to the effective temperature ($T_{\rm eff}$), metallicity ([Fe/H]) and surface gravity ($\log g$) of TrES-3 taken from Sozzetti et al. (2009) [$T_{\rm eff} = 5650$ K, [Fe/H] $= -0.19$, $\log g = 4.4$ (cgs)].

Sozzetti et al. (2009) ($T_{\rm eff} = 5650$ K, [Fe/H] $= -0.19$, $\log g = 4.4$ (cgs)). Table 3 lists the limb darkening coefficients used for each band. Additionally, temporally uncorrelated Gaussian (white) and temporally correlated Gaussian (red) noise were left as free parameters. We accounted for red noise in our analysis because light curves modelled with software that does not account for red noise can provide incorrect best-fitting parameter values and underestimate the errors in those values (Pont, Zucker & Queloz 2006; CW09; Gazak et al. 2011). A thorough description of how the white and red noise are calculated by TAP can be found in CW09 and Gazak et al. (2011).

Pont et al. (2011) found that the upper limit to the eccentricity of TrES-3b is 0.16. We tested if using this eccentricity would change the parameters obtained by TAP on the 2012 April 11 data, and we obtained values within $1\sigma$ of the model using $e = 0$ (see Table 4). Since the parameters derived from TAP do not depend on the upper limit found by Pont et al. (2011) we used an $e = 0$ for this study.

JKTEBOP fits the following parameters: $\frac{R_{\rm p}}{R_*}$, $i$, $T_{\rm c}$, $\mu_1$ and $\mu_2$. It does not fit $\frac{a}{R_*}$, but instead fits the sum of the fractional radii ($\frac{R_{\rm p}}{a} + \frac{R_*}{a}$). In our modelling with JKTEBOP, we allowed $\frac{R_{\rm p}}{R_*}$, $i$, $T_{\rm c}$ and $\frac{R_{\rm p}}{a} + \frac{R_*}{a}$ to float while fixing the limb darkening coefficients to the values listed in Table 3. The eccentricity was also set to zero in the JKTEBOP modelling.

Both JKTEBOP and TAP have been shown to produce the same results in the study of another transiting exoplanet, WASP-5b (Hoyer et al. 2012). However, Hoyer et al. (2012) noticed that JKTEBOP underestimated the errors in the fitted parameters, which they believe was caused by the lack of multi-parameter uncertainty estimation and not accounting for red noise. We used both JKTEBOP and TAP to model our 2012 April 11 data (Table 5) and also found that JKTEBOP underestimated the errors in the fitted parameters. The 2012 April 11 data were chosen for this comparison due its low OoT baseline scatter and a significant amount of red noise (Table 2).

**Table 4.** Parameters derived in this study of the TrES-3b light curves using the TAP.

| Date (UT) | Filter[1] | Midtransit Time (BJD − 245 0000) | $a/R_*$ | $R_p/R_*$ | Inclination (°) | Duration (min) |
|---|---|---|---|---|---|---|
| 2009 June 13 | B | $4995.750\ 25^{+0.00045}_{-0.00051}$ | $5.36^{+0.25}_{-0.18}$ | $0.214^{+0.052}_{-0.039}$ | $79.9^{+1.1}_{-1.0}$ | $87.4^{+1.1}_{-1.1}$ |
| 2009 June 22 | R | $5004.892\ 49^{+0.00020}_{-0.00018}$ | $5.68^{+0.29}_{-0.19}$ | $0.188^{+0.031}_{-0.021}$ | $80.9^{+0.9}_{-0.9}$ | $83.4^{+0.7}_{-0.7}$ |
| 2009 July 04 | V | $5017.954\ 52^{+0.00041}_{-0.00038}$ | $6.02^{+0.32}_{-0.31}$ | $0.166^{+0.023}_{-0.008}$ | $82.0^{+0.6}_{-1.0}$ | $81.7^{+1.0}_{-1.0}$ |
| 2011 October 14 | U | $5848.6873^{+0.0011}_{-0.0011}$ | $7.10^{+1.30}_{-1.10}$ | $0.146^{+0.030}_{-0.015}$ | $84.1^{+2.1}_{-2.4}$ | $75.5^{+2.3}_{-2.3}$ |
| 2011 November 04 | U | $5869.5836^{+0.0017}_{-0.0013}$ | $5.18^{+0.56}_{-1.50}$ | $0.227^{+0.046}_{-0.060}$ | $79.4^{+2.0}_{-1.7}$ | $87.0^{+2.2}_{-2.2}$ |
| 2012 March 25 | U | $6011.959\ 70^{+0.00081}_{-0.00083}$ | $5.85^{+0.43}_{-0.29}$ | $0.208^{+0.066}_{-0.044}$ | $80.9^{+1.5}_{-1.1}$ | $80.0^{+2.3}_{-2.3}$ |
| 2011 March 28(1) | J-C R | $6014.572\ 95^{+0.00073}_{-0.00072}$ | $5.36^{+0.30}_{-0.25}$ | $0.227^{+0.051}_{-0.049}$ | $79.7^{+1.2}_{-1.1}$ | $86.4^{+5.8}_{-5.8}$ |
| 2011 March 28(2) | J-C R | $6014.57287^{+0.00095}_{-0.00089}$ | $6.00^{+1.40}_{-0.56}$ | $0.180^{+0.076}_{-0.028}$ | $81.5^{+2.6}_{-1.8}$ | $77.0^{+5.8}_{-5.8}$ |
| 2012 April 11 | B | $6028.94120^{+0.00049}_{-0.00049}$ | $6.14^{+0.41}_{-0.43}$ | $0.153^{+0.014}_{-0.007}$ | $82.4^{+1.0}_{-1.0}$ | $80.5^{+1.6}_{-1.6}$ |
| 2012 April 11[a] | B | $6028.94066^{+0.00049}_{-0.00047}$ | $5.96^{+0.45}_{-0.37}$ | $0.154^{+0.014}_{-0.008}$ | $81.9^{+1.0}_{-0.9}$ | $81.3^{+1.5}_{-1.5}$ |
| Combined dates | U | — | $5.82^{+0.50}_{-0.27}$ | $0.173^{+0.031}_{-0.018}$ | $81.50^{+1.3}_{-0.9}$ | $81.93^{+0.66}_{-0.66}$ |
| Combined dates | B | — | $5.63^{+0.29}_{-0.19}$ | $0.171^{+0.018}_{-0.013}$ | $81.16^{+0.8}_{-0.6}$ | $84.86^{+0.86}_{-0.86}$ |
| Combined dates | V | — | $6.02^{+0.32}_{-0.31}$ | $0.166^{+0.023}_{-0.008}$ | $81.99^{+0.6}_{-1.0}$ | $81.74^{+0.98}_{-0.98}$ |
| Combined dates | R | — | $5.79^{+0.21}_{-0.21}$ | $0.176^{+0.023}_{-0.011}$ | $81.37^{+0.6}_{-0.5}$ | $83.35^{+0.70}_{-0.70}$ |
| All Dates | — | — | $5.81^{+0.19}_{-0.17}$ | $0.1693^{+0.0087}_{-0.0069}$ | $81.35^{+0.63}_{-0.51}$ | $81.30^{+0.23}_{-0.23}$ |

[1] Filter used: *B* is the Harris B (330–550 nm), *R* is the Harris R (550–900 nm), *V* is the Harris V (473–686 nm), *U* is the Bessell U (303–417 nm) and *J-C R* is the Johnson–Cousins R (523–940 nm).

[a] For this model the eccentricity was fixed at 0.16, which Pont et al. (2011) found to be the upper limit of the eccentricity of TrES-3b.

**Table 5.** Values obtained with JKTEBOP and TAP modelling software with data from the 2012 April 11 transit of TrES-3b.

| Modelling Program | $R_p/R_*$ | Mid-transit Time (HJD − 245 6020) | Inclination (°) | $a/R_*$ | $\frac{R_p}{a} + \frac{R_*}{a}$ |
|---|---|---|---|---|---|
| JKTEBOP | $0.160^{+0.005}_{-0.005}$ | $8.939\ 05^{+0.000\,25}_{-0.000\,25}$ | $81.8^{+0.4}_{-0.4}$ | – | $0.201^{+0.007}_{-0.007}$ |
| TAP | $0.153^{+0.014}_{-0.007}$ | $8.939\ 24^{+0.000\,49}_{-0.000\,49}$ | $82.4^{+0.8}_{-1.0}$ | $6.14^{+0.41}_{-0.43}$ | – |

The $\frac{R_p}{R_*}$, $i$, $T_c$ and $\frac{a}{R_*}$ parameters obtained from TAP and the derived transit durations are summarized in Table 4. We calculated the transit duration, $\tau_t$, of each of our transit model fits with the following equation from Carter et al. (2008):

$$\tau_t = t_{egress} - t_{ingress}, \qquad (2)$$

where $t_{egress}$ is the best-fitting model time of egress, and $t_{ingress}$ is the best-fitting model time of ingress. To calculate the error in the $\tau_t$, we followed Carter et al. (2008) and set the error to twice the cadence of our observations. All nine transits, respective best-fitting models and residuals are illustrated in Fig. 1. All derived parameters are consistent within $1.7\sigma$ among all observed transits, except for $\tau_t$ on the 2009 June 13 ($2.7\sigma$ deviation) and 2011 October 14 ($3.7\sigma$ deviation) transits. The deviation of the 2009 June 13 transit may be caused by starspots, as Lee et al. (2011) and Christiansen et al. (2011) suggested may occur on the surface of TrES-3, since a deviation in $\tau_t$ is not observed in the 2012 April 11 transit taken with the same filter and with worse seeing (Table 2). However, we cannot be certain whether TrES-3 had starspots on 2009 June 13 without conducting a study of at least several years in one photometric bandpass like Zellem et al. (2010) (see their fig. 5). The 2011 October 14 transit is inconsistent with the other transits likely due to technical problems (see Section 2) while obtaining the data since this error occurred right before the transit ingress. The values determined with TAP overlap with previous studies, as shown in Table 6.

Additionally, we performed a combined data analysis for data taken with the same filter and our complete data set. To model each filter, we input all data from the phased light curves and followed the same modelling procedure described above. To model all the phased light curve data in TAP, we followed the procedures described above, except we allowed the limb-darkening coefficients to float instead of setting them at a fixed value. The light curves from the combined data analysis (for filters with multiple observation dates and all data), best-fitting models and residuals are shown in Fig. 2, and the values obtained from TAP are listed in Table 4. Table 7 lists the OoT RMS, residual RMS and red noise for each combined data set (e.g. *U, B, V, R* and all data). In Fig. 3, we show a magnified version of the combined data analysis for the Bessell *U* filter. In this figure, we show the minimum predicted ingress timing difference of 5 min derived in Section 1. It is important to note that the errors in the parameters derived by TAP are higher than most published studies of TrES-3b (see Table 6). We used the errors calculated from TAP which account for red noise using the wavelet likelihood technique (CW09). Therefore, we resulted in more conservative errors than previous studies such as Gibson et al. (2009), despite having similar light curve precision of 1.0–3.3 mmag.

**Table 6.** Literature values of the same parameters derived in this study of TrES-3b's light curves using the TAP.

| Source | $a/R_*$ | $R_p/R_*$ | Inclination (°) | Duration (min) | Period (d) |
|---|---|---|---|---|---|
| O'Donovan et al. 2007 | $6.06^{+0.10}_{-0.10}$ | $0.1660^{+0.0024}_{-0.0024}$ | $82.15^{+0.21}_{-0.21}$ | – | $1.306\ 19^{+0.000\ 01}_{-0.000\ 01}$ |
| Sozzetti et al. 2009 | $5.926^{+0.056}_{-0.056}$ | $0.1655^{+0.0020}_{-0.0020}$ | $81.85^{+0.16}_{-0.16}$ | – | $1.306\ 185\ 81^{+0.000\ 000\ 51}_{-0.000\ 000\ 51}$ |
| Gibson et al. 2009 | – | $0.1664^{+0.0011}_{-0.0018}$ | $81.73^{+0.13}_{-0.04}$ | $79.92^{+1.44}_{-0.60}$ | $1.306\ 1864^{+0.000\ 000\ 5}_{-0.000\ 000\ 5}$ |
| Colón et al. 2010 | – | $0.1662^{+0.0046}_{-0.0048}$ | – | $83.77^{+1.15}_{-2.79}$ | – |
| Southworth 2010 | – | – | $82.07^{+0.17}_{-0.17}$ | – | $1.306\ 1864^{+0.000\ 000\ 5}_{-0.000\ 000\ 5}$ |
| Lee et al. 2011 | – | $0.1603^{+0.0042}_{-0.0042}$ | $81.77^{+0.14}_{-0.14}$ | – | $1.306\ 187\ 00^{+0.000\ 000\ 15}_{-0.000\ 000\ 15}$ |
| Christiansen et al. 2011 | $6.0096^{+0.1226}_{-0.1226}$ | $0.1661^{+0.0343}_{-0.0343}$ | $81.99^{+0.30}_{-0.30}$ | $81.9^{+1.1}_{-1.1}$ | $1.306\ 186\ 08^{+0.000\ 000\ 38}_{-0.000\ 000\ 38}$ |
| Southworth 2011 | – | – | $81.93^{+0.13}_{-0.13}$ | – | $1.306\ 187\ 00^{+0.000\ 000\ 72}_{-0.000\ 000\ 72}$ |
| Sada et al. 2012 | – | – | – | $77.9^{+1.9}_{-1.9}$ | $1.306\ 1865^{+0.000\ 000\ 2}_{-0.000\ 000\ 2}$ |
| This work | $5.81^{+0.19}_{-0.17}$ | $0.1693^{+0.0087}_{-0.0069}$ | $81.35^{+0.63}_{-0.51}$ | $81.30^{+0.23}_{-0.23}$ | $1.306\ 1854^{+0.000\ 000\ 1}_{-0.000\ 000\ 1}$ |

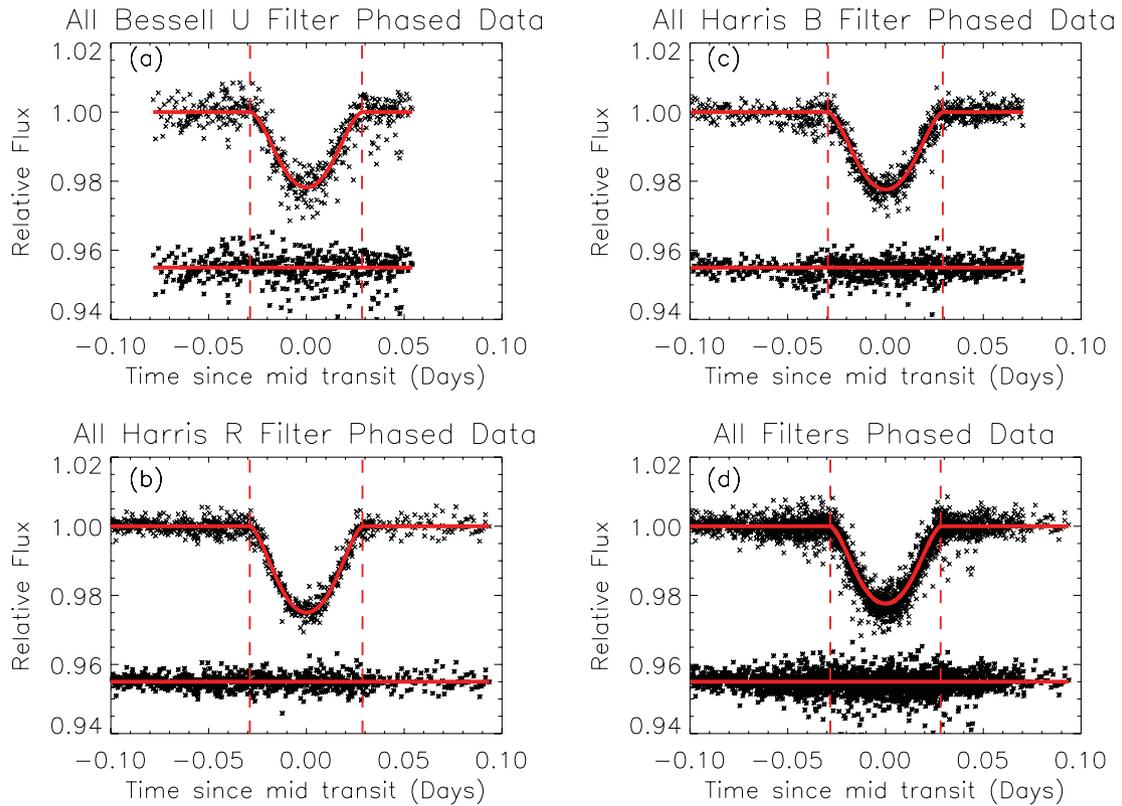

**Figure 2.** Light curves of TrES-3b produced by combining all phased transit data for the (a) Harris B filter, (b) Bessell *U* filter, (c) Harris *R* filter and (d) all filters (see Table 2). The best-fitting models obtained from TAP are shown as a solid red lines. The residuals are shown below each light curve. The TAP best-fitting model predicted ingress and egress points are shown as dashed-red vertical lines. See Table 7 for the OoT RMS flux, residual RMS flux and red noise for each combined light curve.

### 3.2 Period determination

By combining our TAP-derived mid-transit times with previously published mid-transit times of TrES-3b, we can refine the orbital period of the planet and search for any transit timing variations. All mid-transit times ($T_c$) obtained in this paper are compiled in Table 8. When necessary, the literature mid-transit times were transformed from Julian date (JD), which is based on Universal Time Coordinated (UTC) time, into Barycentric Julian Date (BJD), which is based on Barycentric Dynamical Time (TDB), using the online converter[5] by Eastman, Siverd & Gaudi (2010). Using all the mid-transit times in Table 8 and literature values, we derived an improved ephemeris for TrES-3b by performing a weighted linear least-squares analysis using the following equation:

$$T_c = \mathrm{BJD}_{\mathrm{TDB}} T_c(0) + P_b E, \qquad (3)$$

where $P_b$ is the orbital period of TrES-3b and $E$ is the integer number of cycles after the discovery paper (O'Donovan et al. 2007). From this derivation, we obtained values of $T_c(0)$ = 245 4185.912 90 ± 0.000 06 BJD$_{TDB}$ and $P_b$ = 1.306 1854 ± 0.000 0001 d. The periods derived from previous studies are listed in Table 7. We are closest in value to the period derived by Christiansen et al. (2011) of $P_b$ = 1.306 186 08 ± 0.000 000 38 d. The observation minus calculation (O−C) diagram is plotted in Fig. 4 and shows a deviation from our newly derived linear ephemeris. The standard deviation in the O−C plot is ∼2.2 min. Lee et al. (2011) supplemented their observations with amateur astronomy data from the Exoplanet Transit Database[6] (Poddaný, Brát & Pejcha 2010) website and found a larger standard deviation in their O−C plot of 3.6 min. It is possible that the scatter in our O−C plot arises from starspots, as suggested by Lee et al. (2011) and Christiansen et al. (2011). TrES-3 is a G-type star ($T_{eff}$ = 5650 ± 75 K, age = $0.9^{+2.8}_{-0.8}$ Gyr; Sozzetti et al. 2009) and should exhibit starspot behaviour based upon the results of previous studies of main sequence stars (e.g. Wilson 1978, Berdyugina 2005, Strassmeier 2009). Studies of another exoplanet, WASP-4b ($M_b$ = 1.22 $M_{Jup}$, $R_b$ = 1.42 $R_{Jup}$, $a$ = 0.023 au; Wilson 2008), around a G-type star have shown that the loss of light due to starspots can vary between ∼0.1 and 1 per cent depending on their size (Sanchis-Ojeda et al. 2011). Additionally, Sanchis-Ojeda et al. (2011) estimated the shift in the mid-transit time of an exoplanet light curve due to a starspot anomaly (assuming the spot anomaly happened during ingress or egress) to be

$$\Delta t_{spot} \approx 2\,s\left(\frac{A_s}{1.5\,mmag}\right)\left(\frac{T_s}{0.4\,h}\right), \qquad (4)$$

where $A_s$ is the amplitude of the spot anomaly and $T_s$ is the duration of the spot anomaly. Since we do not observe an obvious starspot anomaly, we can estimate what the effect of a starspot within the noise would be. We find a $\Delta t_{spot}$ to be ∼10 s by setting $A_s$ to be twice the precision of the 2009 June 13 transit (Table 2) and $T_s$ of 0.45 h, the average of the starspot anomalies observed by Sanchis-Ojeda et al. (2011). Follow-up observations and more in-depth modelling are needed in order to fully understand if dynamical interactions with other bodies in the system or starspots may be causing this deviation in the $O-C$ plot of TrES-3b.

[5] http://astroutils.astronomy.ohio-state.edu/time/utc2bjd.html

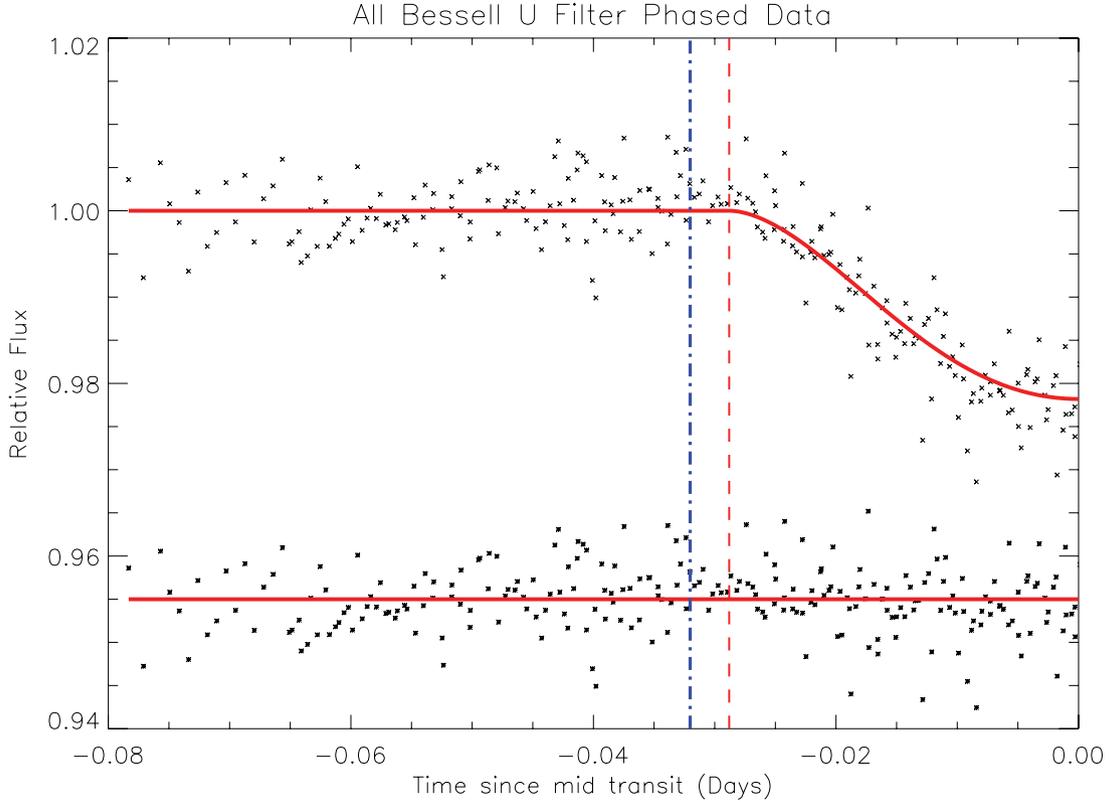

**Figure 3.** The first half of the light curve of TrES-3b produced by combining all phased transit data for the Bessell *U* filter. The dashed–dotted blue line is the minimum timing difference (5 min) between the near-UV and optical ingress found by using a reasonable estimate of $B_* = 100$G and $B_p$ of 8 G (Reiners & Christensen 2010). The TAP best-fitting model predicted ingress is shown as a dashed-red vertical line. Llama et al. (2011) found that the symmetry of a transit light curve is broken when a bow shock is introduced, resulting in an early near-UV ingress and a skewed shape of the transit ingress (the exact shape depends on the plasma temperature, shock geometry and optical depth). We do not see an early transit ingress (assuming a reasonable planetary magnetic field) or an odd shaped transit ingress. The best-fitting model obtained from TAP (assuming spherical symmetry) is shown as the solid red line. The residuals are shown below the light curve.

**Table 7.** Combined light curve properties for the combined data analysis.

| Filter[1] | OoT RMS[2] (mmag) | Res RMS[3] (mmag) | Red noise[4] (mmag) | Average seeing (arcsec) |
|---|---|---|---|---|
| All *U* | 3.72 | 4.10 | $12^{+6}_{-5}$ | 1.56–2.58 |
| All *B* | 2.25 | 2.60 | $12^{+3}_{-3}$ | 1.50–2.49 |
| All *V* | 2.83 | 2.57 | $5.6^{+3.4}_{-3.5}$ | 1.56–2.30 |
| All *R* | 1.69 | 2.00 | $6.5^{+2.8}_{-3.0}$ | 1.49–3.60 |
| All | 2.79 | 2.90 | $6.7^{+3.0}_{-3.1}$ | 1.49–3.60 |

[1] Filter used: *U* is the Bessell U (303–417 nm), *B* is the Harris B (330–550 nm), *V* is the Harris V (473–686 nm) and *R* is both the Harris R (550–900 nm) and Johnson–Cousins R (523–940 nm).
[2] OoT RMS relative flux.
[3] Residual (res) RMS flux after subtracting the TAP best-fitting model from the data.
[4] Red noise (temporally correlated noise) calculated from TAP.

[6] http://var2.astro.cz/ETD/

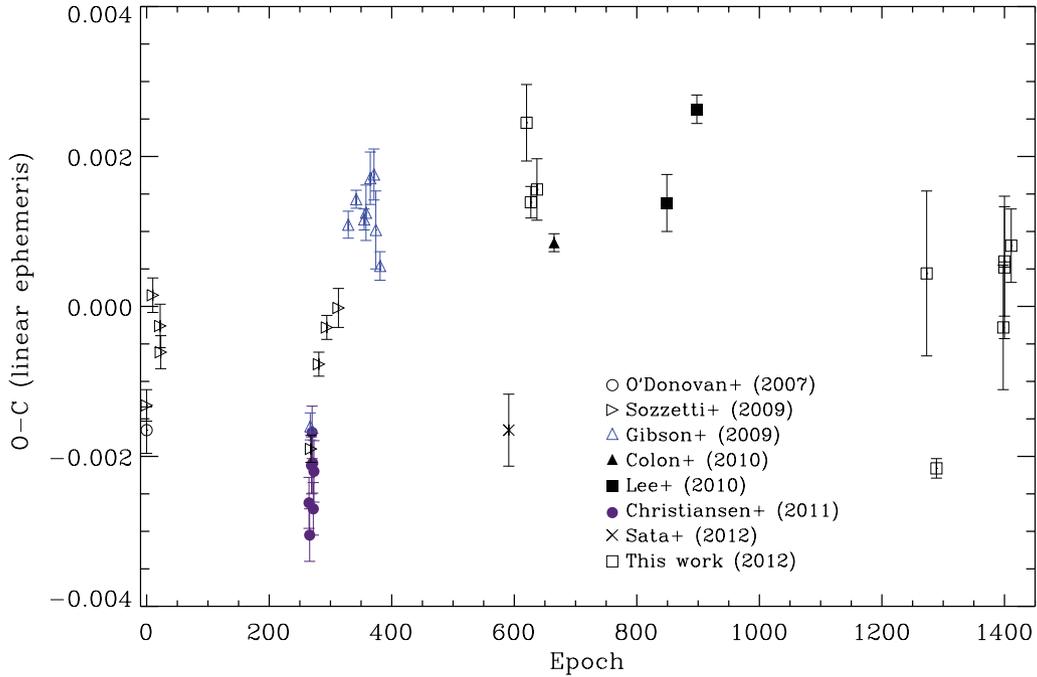

**Figure 4.** The observation minus calculation ($O-C$) diagram for TrES-3b produced from the linear ephemeris calculated in this paper. This plot can help determine if TrES-3b is exhibiting transit timing variations (TTVs) due to other bodies in the system or starspots on its host star. The standard deviation in this plot is ∼2.2 min, which may be caused by starspots as suggested by Lee et al. (2011) and Christiansen et al. (2011). No obvious TTVs are seen, however, more in-depth dynamical modelling is needed to fully understand if dynamical interactions with other bodies in the system may be causing this deviation in the $O-C$ plot of TrES-3b. The times given by each reference are shown as different shapes, and the times derived in this study are marked as black open boxes. See Table 8 for the values that are plotted from this study.

**Table 8.** Results of the transit timing analysis.

| Source | Date (UT) | $T_c$ (BJD$_{TDB}$) | $T_c$ error (d) | Epoch | $O-C$ (d) | $O-C$ error (d) |
|---|---|---|---|---|---|---|
| This paper | 2009 June 13 | 245 499 5.750 25 | 0.000 51 | 620 | 0.002 45 | 0.000 51 |
| | 2009 June 22 | 245 500 4.892 49 | 0.000 20 | 627 | 0.001 39 | 0.000 21 |
| | 2009 July 04 | 245 501 7.954 52 | 0.000 41 | 637 | 0.001 56 | 0.000 41 |
| | 2011 October 14 | 245 584 8.687 3 | 0.001 1 | 1273 | 0.000 4 | 0.0011 |
| | 2011 November 04 | 245 586 9.583 63 | 0.000 12 | 1289 | −0.002 16 | 0.000 13 |
| | 2012 March 25 | 245 601 1.959 70 | 0.000 83 | 1398 | −0.000 28 | 0.000 83 |
| | 2012 March 28(1) | 245 601 4.572 95 | 0.000 73 | 1400 | 0.000 60 | 0.000 73 |
| | 2012 March 28(2) | 245 601 4.572 87 | 0.000 95 | 1400 | 0.000 52 | 0.000 95 |
| | 2012 April 11 | 245 602 8.941 20 | 0.000 49 | 1411 | 0.000 81 | 0.000 49 |

## 4 DISCUSSION

### 4.1 Searching for asymmetries between the near-UV and optical light curves

VJH11a predicted that TrES-3b should exhibit an early near-UV ingress. This effect has been observed in WASP-12b with five *Hubble Space Telescope* spectroscopic data points using the Cosmic Origins Spectrograph in the NUVA (253.9–258.0 nm) near-UV filter (FHF10). FHF10 found that the near-UV light curve of WASP-12b started approximately 25–30 min earlier than its optical light curve, and also exhibited a near-UV $R_p/R_*$ ∼0.1 greater than in the optical (FHF10, see Fig. 2). We do not observe such a large early ingress in the near-UV light curve compared to the optical light curve or a significant $R_p/R_*$ difference in the transits of TrES-3b. As seen in Table 1, the planetary parameters ($M_b$, $R_b$, $P_b$, a) of WASP-12b and TrES-3b are very similar, the stellar parameters ($R_*$, $M_*$, [Fe/H]) of WASP-12 and TrES-3 are slightly different and the WASP-12 $B_p/B_*$ value calculated from VJH11a is ∼6.8 times higher than the TrES-3 system. In Section 1, we found a reasonable timing difference between the near-UV and optical ingress of 5–11 min for $B_* = 100$ G and $B_p$ ranging from 8–30 G. By combining this prediction with a higher $B_p/B_*$ for the WASP-12b system, it is not surprising that we do not see a 25–30 min early ingress in TrES-3b.

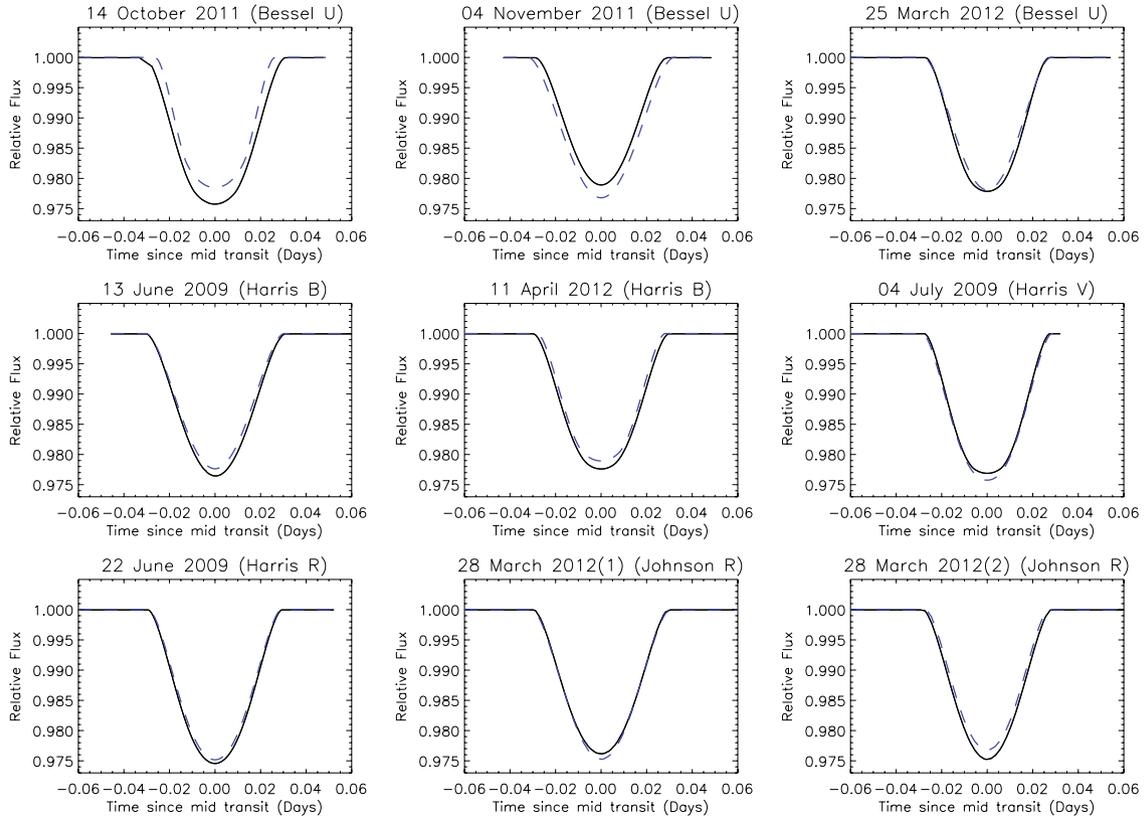

**Figure 5.** Each panel shows the best-fitting models obtained from the ᴛᴀᴘ (dashed-blue line; Fig. 1 shows the goodness of fit of the data by ᴛᴀᴘ) plotted with those obtained by ᴊᴋᴛᴇʙᴏᴘ (solid black line) for each observation and filter. None of the transits show any significant non-spherical asymmetries between models except 2011 October 14, which is likely due to technical problems (Section 2) when taking the data. We obtain an RMS value between 0.05 and 0.20 mmag (which is below the precision of every observation; Table 2) by subtracting the ᴊᴋᴛᴇʙᴏᴘ and ᴛᴀᴘ models from each other for all dates. This result implies that any spherical differences seen between models (e.g. 2011 November 04) is below the precision of our light curves.

In an attempt to detect much smaller differences in the near-UV ingress (of the order of several minutes) and $R_p/R_*$ (on the order of ∼0.05) in the light curves of TrES-3b, we carefully examined our data and different models for signs of transit asymmetry. Fig. 5 shows the ᴛᴀᴘ and ᴊᴋᴛᴇʙᴏᴘ best-fitting models, which overlap. The only transit that shows a non-spherical asymmetry is the 2011 October 14 transit using the Bessell *U* filter. This discrepancy is likely caused by the technical problems (see Section 2) at the start of the transit and the lack of data points during that time. By subtracting the ᴛᴀᴘ best-fitting models from the ᴊᴋᴛᴇʙᴏᴘ best-fitting models, we obtain an RMS value of 0.05–0.20 mmag, which is below the precision of all the respective light curves. Therefore, any spherical asymmetries between models (e.g. 2011 November 04) cannot be trusted. From these results, it is clear that the near-UV transits do not display any non-spherical asymmetries since the ᴊᴋᴛᴇʙᴏᴘ (which is capable of accounting for non-spherical asymmetries) and ᴛᴀᴘ (which assumes the planet is spherical) models are nearly identical. These findings imply that TrES-3b is spherical in all wavelength bands. Additionally, Llama et al. (2011) presented models of the bow shock of WASP-12b, all with an obvious asymmetry between the two halves of the transit (Llama et al. 2011, see fig. 2). Following this idea, we subtracted each light curve by the mirror image of itself about the ᴛᴀᴘ-calculated mid-transit time. We used this technique to find possible asymmetries on either half of the transit. From this test the RMS values for each date range between 0.20 and 0.40 mmag, except for the 2011 October 14 transit (which has an RMS of 1.1 mmag). These RMS values are all below the precision of each respective transit (see Table 2). The results of this experiment indicate no asymmetries above the noise levels, and imply that all the transits of TrES-3b are spherically symmetric. As shown in Fig. 3, we should have seen an early ingress of at least 5 min (assuming a reasonable $B_p$ and $B_*$) with the timing resolution and precision of our near-UV light curve. We cannot compare our near-UV transit shapes with a predicted transit shape using VJH11a's model because Llama et al. (2011) found that the symmetry of a transit light curve is broken when a bow shock is introduced resulting in the shape of the transit ingress being skewed (the exact shape depends on the plasma temperature, shock geometry and optical depth). Lastly, we find an average near-UV (all U-band data) $R_p/R_*$ of $0.173^{+0.031}_{-0.018}$ and an average optical (all *V*-band data) $R_p/R_*$ of $0.166^{+0.023}_{-0.008}$, which are not statistically different from each other. We conclude that all near-UV and optical light curves of TrES-3b are symmetrical and do not show any asymmetries between each other.

Even though we do not observe a timing difference between our near-UV and optical light curves, it is still possible to calculate an upper limit to the magnetic field of TrES-3b. The near-UV light curves of TrES-3b might contain a timing difference from the optical light curves, but it is just below the cadence sampled in our transits. The Nyquist–Shannon sampling theorem states that a signal has to be sampled at the Nyquist frequency (with a period equal to two sampling intervals) in order to fully detect the signal. Therefore, we use twice the 2012 March 25 near-UV light curve cadence (since it has the highest near-UV transit cadence; Table 2) for the input of the timing difference into equation (1). Any timing difference below this cannot be detected in our data set. Our cadence is longer than the minimum timing difference predicted by VJH11a of 3 s and shorter than the cadence achieved for the WASP-12b observations of 5733 s (FHF10). Using a timing difference of 138 s, we determine an upper limit on the magnetic field of TrES-3b to be 0.013 $B_*$. For TrES-3's magnetic field strength, we used a range 1–100 G (consistent with studies of G-type stars; Plachinda & Tarasova 2000, 1999; Plachinda 2004; Reiners 2012) to find a range for the upper limit on TrES-3b's magnetic field to be 0.13–1.3 G. This upper limit is unexpected for the age of the TrES-3 system because the predicted magnetic field strengths for 1 $M_{Jup}$ exoplanets range from 8 G (Reiners & Christensen 2010) to 30 G (Sánchez-Lavega 2004; Reiners & Christensen 2010).

**Table 9.** Physical Properties of the TrES-3 system derived from the light curve modelling.

| Filter[1] | $M_b$ ($M_{Jup}$) | $R_b$ ($R_{Jup}$) | $\rho_b$ ($\rho_{Jup}$) | log $g_b$ (cgs) | $T'_{eq}$ (K) | $\Theta$ | $a$ (au) |
|---|---|---|---|---|---|---|---|
| All $U$ | $1.904^{+0.087}_{-0.087}$ | $1.386^{+0.248}_{-0.144}$ | $0.67^{+0.48}_{-0.28}$ | $3.17^{+0.69}_{-0.53}$ | $1656^{+40}_{-52}$ | $0.066^{+0.013}_{-0.008}$ | $0.0223^{+0.0019}_{-0.0010}$ |
| All $B$ | $1.906^{+0.055}_{-0.055}$ | $1.370^{+0.192}_{-0.104}$ | $0.69^{+0.39}_{-0.21}$ | $3.19^{+0.52}_{-0.53}$ | $1684^{+29}_{-35}$ | $0.065^{+0.010}_{-0.005}$ | $0.0216^{+0.0011}_{-0.0007}$ |
| All $V$ | $1.902^{+0.067}_{-0.067}$ | $1.330^{+0.184}_{-0.064}$ | $0.76^{+0.42}_{-0.15}$ | $3.21^{+0.43}_{-0.61}$ | $1628^{+44}_{-41}$ | $0.071^{+0.011}_{-0.005}$ | $0.0230^{+0.0013}_{-0.0012}$ |
| All $R$ | $1.905^{+0.062}_{-0.062}$ | $1.410^{+0.188}_{-0.088}$ | $0.64^{+0.33}_{-0.16}$ | $3.01^{+0.21}_{-0.34}$ | $1660^{+46}_{-33}$ | $0.065^{+0.009}_{-0.005}$ | $0.0222^{+0.0007}_{-0.0007}$ |
| All | $1.905^{+0.083}_{-0.083}$ | $1.356^{+0.070}_{-0.055}$ | $0.71^{+0.15}_{-0.12}$ | $3.15^{+0.38}_{-0.30}$ | $1657^{+31}_{-29}$ | $0.068^{+0.004}_{-0.004}$ | $0.0222^{+0.0008}_{-0.0008}$ |

[1]Filter used: $U$ is the Bessell U (303–417 nm), $B$ is the Harris B (330–550 nm), $V$ is the Harris V (473–686 nm) and $R$ is both the Harris R (550–900 nm) and Johnson–Cousins R (523–940 nm).

Our result implies that either the magnetic field of TrES-3b is abnormally low, the VJH11a method to determine exoplanet magnetic fields cannot be applied to TrES-3b, or the effect proposed by VJH11a can only be observed with near-UV wavelengths not accessible from the ground (Turner et al. 2012b). In addition, our result may also suggest that the techniques outlined by VJH11a can only be used in narrow-band spectroscopy (as with the WASP-12b observations) not broad-band photometry. However, a full radiative transfer analysis is needed to verify this possibility. Additionally, the spectral region (253.9–258.0 nm) covered by early near-UV observations of WASP-12b with the *HST* (FHF10) includes resonance lines from Na I, Al I, Sc II, Mn II, Fe I, Co I and Mg I (Morton 1991, 2000) and these species also have strong spectral lines in our $U$ band (303–417 nm; Sansonetti, Martin & Young 2005). Alternatively, the effect proposed by VJH11a may be more dependent on the metallically of the host star than the near-UV wavelength region observed. TrES-3 has a metallicity of $-0.19^{+0.08}_{-0.08}$ (Sozzetti et al. 2009), which is significantly lower than WASP-12's metallicity of $0.30^{+0.05}_{-0.15}$ (Hebb et al. 2009). Obtaining more observations of other exoplanets predicted by VJH11a to exhibit an early near-UV ingress will help distinguish between all the possibilities discussed above (Turner et al. in preparation; Turner et al. 2012a, b; Walker-LaFollette et al. 2012). Future observations with other telescopes capable of achieving a better near-UV cadence are needed to verify our conclusions, constrain VJH11a's techniques and to search for bow shock temporal variations predicted by Vidotto et al. (2011c). Furthermore, Rossiter–McLaughlin effect (Winn 2011) measurements are needed to constrain possible bow shock variability by determining whether TrES-3b is in the corotation radius of its host star. We also advocate for low-frequency radio emission and magnetic star–planet interaction observations of TrES-3b to further constrain its magnetic field and to supplement our findings.

## 4.2 Physical properties of the TrES-3 system

We used the results of our light curve modelling to calculate planetary and geometrical parameters of TrES-3b, including its mass, radius, density, surface gravity, equilibrium temperature, Safronov number and semi-major axis. Specifically, the TAP values derived from our combined data analysis for our entire data set and each filter (Table 4) were used to derive the desired parameters.

We adopted the formula by Southworth, Wheatley & Sams (2007) to calculate the surface gravitational acceleration, $g_b$:

$$g_b = \frac{2\pi}{P_b} \left(\frac{a}{R_b}\right)^2 \frac{\sqrt{1-e^2}}{\sin i} K_*,$$ (5)

where $K_*$ is the stellar velocity amplitude from Sozzetti et al. (2009), listed in Table 9.

The equilibrium temperature, $T_{eq}$, was derived using the relation (Southworth 2010)

$$T_{eq} = T_{eff} \left(\frac{1-A}{4F}\right)^{1/4} \left(\frac{R_*}{2a}\right)^{1/2},$$ (6)

where $T_{eff}$ is the effective temperature of the host star at 5650 K (Sozzetti et al. 2009), $A$ is the Bond albedo and $F$ is the heat redistribution factor. This formula is simplified by making the assumption, as in Southworth (2010), that $A = 1 - F$; the resulting equation is the modified equilibrium temperature, $T'_{eq}$:

$$T'_{eq} = T_{eff} \left(\frac{1}{4}\right)^{1/4} \left(\frac{R_*}{2a}\right)^{1/2}.$$ (7)

The planetary mass, $M_b$, can be calculated using the following equation derived from Seager (2011):

$$M_b = \left(\frac{1}{28.4329}\right) \left(\frac{\gamma}{\sin i}\right) \left(\frac{P_b}{1\,yr}\right)^{1/3} \left(\frac{M_*}{M_\odot}\right)^{2/3} M_{Jup},$$ (8)

where $\gamma$ is the radial velocity semi-amplitude equal to 369.8 ± 7.1 m s$^{-1}$ (Sozzetti et al. 2009) and we use $M_* = 0.921^{+0.014}_{-0.014}$ M$_\odot$ (Southworth 2011).

We calculated the Safronov number, $\Theta$, using the equation from Southworth (2010):

$$\Theta = \frac{M_b a}{M_* R_b}.$$ (9)

The Safronov number is a measure of the ability of a planet to gravitationally scatter other bodies (Safronov 1972). As defined by Hansen & Barman (2007), Class I hot Jupiters have $\Theta = 0.07 \pm 0.01$ and Class II have $\Theta = 0.04 \pm 0.01$.

Results of the $M_b$, $R_b$, the planetary density ($\rho_b$), log $g_b$, $T'_{eq}$, $\Theta$ and $a$ from our analysis are summarized in Table 9. We used the values for the physical parameters of the TrES-3 system by Southworth (2011) to derive the physical properties in Table 9. For all the planetary parameters, our results are within $1\sigma$ of the published literature values. We find a near-UV radius of $R_p = 1.386^{+0.248}_{-0.144}\, R_{Jup}$, which is consistent with the optical radius of $1.310 \pm 0.019\, R_{Jup}$ (Southworth 2011) and therefore, we do not observe a wavelength dependence in the radius of TrES-3b through our analysis. Combining this result with previous optical and near-IR studies, TrES-3b appears to have a constant planetary radius in near-UV through near-IR wavelengths (de Mooij & Snellen 2009 used the William Herschel Telescope on La Palmato to find a near-IR planetary radius of $1.338 \pm 0.019\, R_{Jup}$) accessible from the ground, suggesting a nearly flat atmospheric spectrum in these wavelength bands. Our result is consistent with other transiting exoplanet observations of a flat spectrum from visible to near-IR wavelengths on HD 189733b (a hot Jupiter; Grillmair 2007) and GJ 1214b (a super-Earth; Bean et al. 2011). Fortney et al. (2006) has shown that an isothermal pressure–temperature profile may suppress day-side spectral features (as might be the case for HD 189733b). Nonetheless, an in-depth radiative transfer model needs to be done to fully understand what may be causing the flat atmospheric spectra of TrES-3b.

## 5  CONCLUSIONS

We have investigated nine primary transits of TrES-3b observed between 2009 and 2012 in several optical and near-UV filters. In this study, we derived a new set of planetary system parameters ($M_b = 1.905^{+0.083}_{-0.083}\, M_{Jup}$, $R_b = 1.356^{+0.070}_{-0.055}\, R_{Jup}$, $\rho_b = 0.71^{+0.15}_{-0.12}\, \rho_{Jup}$, $\log g_b = 3.15^{+0.03}_{-0.30}$, $T'_{eq} = 1657^{+31}_{-29}$ K, $\Theta = 0.068^{+0.004}_{-0.004}$, $a = 0.0222^{+0.0008}_{-0.0008}$ au). We find that TrES-3b's near-UV planetary radius of $R_p = 1.386^{+0.248}_{-0.144}\, R_{Jup}$ is consistent within the error of its optical radius of $1.310 \pm 0.019\, R_{Jup}$ (Southworth 2011). Additionally, we updated the orbital period to $P_b = 1.306\ 1854 \pm 0.000\ 0001$ d, and we present an improved ephemeris of $T_c = $ BJD$_{TDB}$245 4185.912 90 $\pm$ 0.000 06 $+$ 1.306 1854$\pm$0.000 0001$E$. Our data include the only published near-UV light curve of TrES-3b. Combining our results with previous near-IR studies (de Mooij & Snellen 2009), TrES-3b appears to have a constant planetary radius in near-UV through near-IR wavelengths.

We did not detect an early near-UV ingress as proposed by VJH11a, despite a detectable timing difference range of 5–11 min (for $B_* = 100$ G and $B_p$ ranging 8–30 G). Our near-UV observations were stable over a 6-month period. Despite this non-detection, we are still able to find an upper limit to the magnetic field of TrES-3b. Since we cannot distinguish a difference in the timing between the light curves below the Nyquist frequency, we use a timing difference of twice the maximum near-UV light curve cadence (138 s) to derive a range on the upper limit of TrES-3b's magnetic field to be 0.013–1.3 G (for a 1–100 G range for TrES-3's magnetic field strength). This upper limit is in contrast to the predicted magnetic field strengths for 1 $M_{Jup}$ exoplanets for the estimated age of the TrES-3 system range between 8 G (Reiners & Christensen 2010) and 30 G (Sánchez-Lavega 2004; Reiners & Christensen 2010). Due to this result, we advocate for follow-up studies on the magnetic field of TrES-3b using other detection methods (such as radio emission and magnetic star–planet interactions) and other telescopes capable of achieving a better near-UV cadence to verify our findings, the techniques of VJH11a and to search for bow shock temporal variations predicted by Vidotto et al. (2011c). Our findings imply that the magnetic field of TrES-3b is abnormally low, that the VJH11a method to determine exoplanet magnetic fields cannot be applied to TrES-3b (possibly due to the low metallicity of its host star) or the effect proposed by VJH11a can only be observed with near-UV wavelengths not accessible from the ground. To help further distinguish between these possibilities, we also encourage observations of other exoplanets predicted by VJH11a to exhibit an early near-UV ingress. Finally, an in-depth radiative transfer analysis is needed to determine whether VJH11a's techniques can only be used in narrow-band spectroscopy and not broad-band photometry.

## ACKNOWLEDGMENTS

We sincerely thank the University of Arizona Astronomy Club, the Steward Observatory TAC, the Steward Observatory telescope day crew, Dr Elizabeth Green, Dr Peter Milne, Dr John Bieging, Dr Serena Kim, Dr Hwankyung Sung, Mr Bumdu Lim, Ms Maria Schuchardt, the Lunar and Planetary Laboratory, the Catalina Sky Survey, the Associated Students of the University of Arizona, Robert Thompson, Juan Lora and Johanna Teske for supporting this research. We would also like to thank the anonymous referee for their insightful comments during the publication process. This manuscript is much improved thanks to their comments.

This paper has been typeset from a TeX/LaTeX file prepared by the author.